\documentclass[
  journal=terrence,
  manuscript=research-paper, 
  year=2020,
  volume=37,
]{cup-journal}

\usepackage[colorlinks,citecolor=blue,urlcolor=blue,bookmarks=false,hypertexnames=true]{hyperref} 
\usepackage{dirtytalk}
\usepackage{adjustbox}
\usepackage{microtype,siunitx,booktabs}
\usepackage{amsmath}
\usepackage{subcaption}

\title{Joint Study of Above Ground Biomass and Soil Organic Carbon for Total Carbon Estimation using Satellite Imagery in Scotland}

\author{Terrence Chan, Carla Arus Gomez, Anish Kothikar, Pedro Baiz}
\affiliation{Imperial College London}
\email[Terrence Chan, Pedro Baiz]{ckc116@imperial.ac.uk, p.m.baiz@imperial.ac.uk}

\doi{10.1017/pasa.2020.32}

\published{8 May 2022}


\begin{document}
\setcitestyle{square}

\begin{abstract}

Land Carbon verification has long been a challenge in the carbon credit market. Carbon verification methods currently available are expensive, and may generate low-quality credit. Scalable and accurate remote sensing techniques enable new approaches to monitor changes in Above Ground Biomass (AGB) and Soil Organic Carbon (SOC). The majority of state-of-the-art research employs remote sensing on AGB and SOC separately, although some studies indicate a positive correlation between the two. We intend to combine the two domains in our research to improve state-of-the-art total carbon estimation and to provide insight into the voluntary carbon trading market. We begin by establishing baseline model in our study area in Scotland, using state-of-the-art methodologies in the SOC and AGB domains. The effects of feature engineering techniques such as variance inflation factor and feature selection on machine learning models are then investigated. This is extended by combining predictor variables from the two domains. Finally, we leverage the possible correlation between AGB and SOC to establish a relationship between the two and propose novel models in an attempt outperform the state-of-the-art results. We compared three machine learning techniques, boosted regression tree, random forest, and xgboost. These techniques have been demonstrated to be the most effective in both domains.
\newline
This research makes three contributions: i) Including Digital Elevation Map (DEM) as a predictor variable in the AGB model improves the model result by 13.5 \% on average across the three machine learning techniques experimented, implying that DEM should be considered for AGB estimation as well, despite the fact that it has previously been used exclusively for SOC estimation. ii) Using SOC and SOC Density improves the prediction of the AGB model by a significant 14.2\% on average compared to the state-of-the-art baseline, which strengthens our experiment results and suggests a future research direction of combining AGB and SOC as a joint study domain. iii) Including AGB as a predictor variable for SOC improves model performance for Random Forest, but reduced performance for Boosted Regression tree and XG Boost, indicating that the results are specific to ML models and more research is required on the feature space and modelling techniques.
Additionally, we propose a method for estimating total carbon using data from Sentinel 1, Sentinel 2, Landsat 8, Digital Elevation, and the Forest Inventory.
\end{abstract}

\section{INTRODUCTION}
\label{sec:int}

\subsection{Climate Urgency and Voluntary Carbon Markets}
\indent 
Global efforts to reduce carbon emissions have gained enormous traction in recent years, owing to the growing urgency of combating climate change \cite{IPCC_press}. The Paris agreement \citep{paris_climate_agreement}, which was adopted by 196 Parties at COP21 \citep{cop21}, outlines the most extensive international collaboration, with the goal to limit global warming to below 2 degree Celsius above pre-industrial levels. Since then, an exploding amount of organisations have emerged to aid in the fight against climate change. Apart from government efforts, the private sector contributes significantly to carbon emission reduction. TSVCM \cite{tsvcm} developed a blueprint for establishing a large-scale transparent carbon trading market. Carbon trading enables businesses to finance carbon sequestration projects in exchange for carbon credits, thereby speeding up the process of carbon removal from the atmosphere.
\newline \indent 
There are three main scopes in the voluntary carbon market, namely supply, market infrastructure and demand \citep{tsvcm1}. Our project aims to contribute to the supply scope by developing reliable and scalable carbon verification methods.
The most critical factor affecting the supply of carbon credits is quality: 
\say{i) Quality of individual projects measured against independent standards and ii) perceptions of offsetting in catalyzing progress toward decarbonization.} \citep{tsvcm1}. 
\newline\indent To ensure adequate supply, early carbon credit suppliers used their own standards to determine the amount of carbon emissions that a project could offset. However, some standards were later discovered to be unreliable, jeopardising a project's credibility. Carbon credit verification can be costly, and the market is currently oversupplied with low-quality credits. The fundamental issue remains how to effectively verify the amount of carbon sequestered by a project over its lifetime, and we hope to shed light on this subject by examining carbon estimation methods in soil and forests.
\newline \indent
Agriculture, forestry, and other land use activities contribute significantly to global CO2 emissions \citep{epa_emissions}. By gaining a better understanding of how soil and forests act as carbon sinks, there is a great deal of potential to motivate land-based carbon sequestration projects. Traditionally, carbon was quantified through fieldwork and laboratory analysis. It requires a huge amount of labour and is regional in nature. Numerous studies used extrapolation methods from localised data to create large scale maps. Additionally, as the space industry has grown, the cost of satellite imagery has decreased significantly, resulting in a burst of interest in remote sensing techniques for carbon measurements.

\subsection{Above Ground Biomass (AGB)}
Above Ground biomass refers to \say{All biomass in living vegetation, both woody and
herbaceous, above the soil including stems, stumps, branches, bark, seeds and foliage.} \citep{Ravindranath2008}.
Among the diverse types of biomass, the majority of research is focused on forests due to their significance in carbon storage. Numerous studies have been conducted on the mapping of AGB stocks in forests, most of which are extremely successful at predicting carbon stocks. \textbf{Biome Averages} is a baseline method that gives a rough estimate on forest carbon stocks \citep{Gibbs_2007}. We estimate an average carbon value according to forest category. This gives us a generalised estimation with high uncertainty but with very low cost. \textbf{Forest Inventory} enhances the baseline by examining the relationship between tree diameters/volumes and carbon stocks. This is a very direct method that is expensive due to the high cost of field labour, limiting its scalability.
\newline \indent
When measuring AGB, the objective is to map large areas of forests while maintaining good accuracy. It is nearly impossible to directly measure each tree using conventional methods. With remote sensing, we can map large areas at a low cost. \textbf{Radar} is extensively used in AGB mapping \citep{Goetz2009} \citep{KASISCHKE1997141} in the past two decades. It measures microwave reflections to estimate vertical structures of trees. \textbf{Optical Remote Sensing (Optical RS)} measures visible and infra-red wavelengths, which produces carbon predictions together with field measurements. Mostly obtained from satellites, Optical RS images are freely available and collected at intervals of days. However, uncertainty is high due to limitations in spectral and spatial resolutions, there are also challenges associated with cloud coverage and other climate conditions. \textbf{Very High-Resolution Airborne Optical RS} captures high-resolution images (0.1-0.2 m) to predict allometry between tree height, crown and carbon stocks. The enhanced spectral resolution allows us to capture the chemical compositions of tree canopy and tree species, which is valuable in carbon prediction. However, the cost is high which limits scalability. \textbf{Light Detection and Ranging (LiDAR)} are very successful in approximating spatial dimensions of forests and are widely used in the modelling of Top of Canopy Height models (TCH). Spaceborne LiDAR cap map global vertical tree structures \citep{Spaceborne_lidar}, together with the incorporation of ML models, can be very accurate. Lang et al. \cite{DBLP:journals/corr/abs-2103-03975} created an accurate supervised CNN model to interpret Global Canopy Height with data sourced from GEDI (NASA's Global Ecosystem Dynamics Investigation). TCH models are very important in the state of the art biomass mapping as it provides information about the 3D structure of forests. There is a wide range of literature mapping biomass carbon content using a mixture of RS techniques with optical sensors providing the 2D area map and TCH models providing the 3D structure for the biomass mapping model \citep{tch2biomass} \citep{tch2biomass2} \citep{tch2biomass3}.
\newline \indent
A summary of literatures on AGB is presented in table \ref{table:agb_literatures}.
The majority of AGB literature is concerned with forest ecosystems throughout the world, with techniques ranging from Linear Regression, Random Forest, and XGBoost \citep{li_2020} \citep{Li2019} to Support Vector Machine \citep{rapideye_2019}. Most literature make use of freely available satellite imagery at the time of the study, while some make use of high-resolution commercial images \citep{LIDAR201801}. In general, none of the AGB research looked into using Digital Elevation as a predictor and none jointly studied AGB estimation with SOC.

\begin{table}
\centering
\begin{adjustbox}{width=1\textwidth,center=\textwidth}
\begin{tabular}{ | m{5em} | m{10em} | m{8em} | m{5em} | m{8em} | m{5em} | m{4em} | m{4em} |} 
\hline
\textbf{Literature} & \textbf{Description} & \textbf{ML Techniques} & \textbf{Vegetation Cover} & \textbf{Data Sources} & \textbf{Region/Year of Study} & \textbf{Use Digital Elevation} & \textbf{Use SOC as a predictor} \\
\hline
Li et al. \cite{li_2020} 
& Compared performances of LR, RF, XGB on AGB prediction in Forest Ecosystems in China. Created novel method to predict AGB with a combination of satellite images. Successfully achieved R$^2$ score of 0.75 ML
& LR, RF, XGBoost; Achieved performance of R$^2$ = 0.75, RMSE = 18.92 MgC/ha (Combined model, XGB)
& Forest Ecosystem 
& LandSat 8, Sentinel 1A, China National Forest Continuous Inventory (NFCI)
& Chenzhou, Hunan, China, 2020 
& No & No \\
\hline
Li et al. \cite{Li2019}
& Explored the influence of different variables and forest types on AGB prediction. Proved that LR is not general enough to model AGB and is outperformed by RF and XGB 
& LR, RF, XGBoost; Achieved performance of R$^2$ = 0.71, RMSE = 24.02 MgC/ha (XGB on all forest types)
& Forest Ecosystem
& LandSat 8, China National Forest Continuous Inventory (NFCI)
& Hunan, China, 2019 & No & No \\ 
\hline
Hojas Gascon et al.\cite{rapideye_2019}
& Estimated AGB using high resolution satellite data (5m) to reduce cost of using field measurements for AGB Estimation. Analysed relationship between AGB from field plots and RS biophysical parameters
& SVM, RF; Achieved performance of R$^2$ = 0.69 on RF model
& Miombo woodland, heavily disturbed coastal forests, arid shrub-lands, and dense forests 
& RapidEye
& Tanzania, East Coast of Africa, 2019 & No & No \\
\hline
Hirata et al. \cite{LIDAR201801} 
& Developed a method for developing countries to estimate AGB using LiDAR and Quick Bird satellite. First performed classification to identify target woodland areas then apply multi-regression model for AGB estimation 
& LR; R$^2$ = 0.73 using satellite data
& Forest Ecosystem
& LiDAR, Quick Bird very high resolution satellite images (0.61/2.44m$^2$ per pixel)
& Kampong Thom Province, central Cambodia, 2018 
& No & No \\
\hline
Asner et al. \cite{malay_2018}
& Combined LiDAR with satellite imagery and other geospatial data to map forest carbon aboveground density with a spatial resolution of 30m$^2$ per pixel. Found unlogged forest contains large amount of carbon and provided guidance on new area for forest protection in Borneo
& RF; R$^2$ = 0.7, RMSE = 41.6 MgC/ha
& Forest Ecosystem
& LiDAR, LandSat 8, Sentinel 1
& Malaysian Borneo, 2018
& No & No \\
\hline

\end{tabular}
\end{adjustbox}
\caption{Summary remote sensing and AGB literatures}
\label{table:agb_literatures}
\end{table}

\subsection{Soil Organic Carbon (SOC)}
Soil Organic Matter makes up around 2 - 10 \% of the total soil mass \citep{som_ausgov}. We are mostly interested in Soil Organic Carbon (SOC), which is the measurable component of Soil Organic Matter.
Globally, soil contains approximately 2344 gigatons of organic carbon \citep{STOCKMANN201380}, which is over 80\% of all terrestrial organic carbon \citep{global_carbon_distribution}, forming the largest terrestrial organic carbon pool. This number is also over 2.7 times compared to atmospheric carbon \citep{global_carbon_distribution}.
\newline \indent
Two quantities are required to calculated SOC:
\textbf{Bulk Density} is \say{ratio of the mass to the bulk or macroscopic volume of soil particles plus pore spaces in a sample} \citep{bulk_density}. It is usually obtained from drying soil samples and obtaining the weight per hectare (kg/ha). \textbf{\% SOC} is found using lab analysis such as Walkey and Black \citep{walkley}. 
\textbf{SOC} is the multiplication of the two quantities and is measured in t C/ha (tonnes of carbon per hectare) \citep{soc_measure_ausgov}.
\newline \indent
Recent literature in the domain of SOC estimation is presented in table \ref{table:soc_literature}. There is a mix of research predominantly on forests and cultivated land using SVM, BRT and RF \citep{zhou_2020} \citep{zhou_2021}. There is also interest in using neural networks to estimate SOC content for mixed land \citep{nortern_iran_2020_soc}. A wide range of satellite data is used including Sentinel 1,2,3, Landsat 8 and MODIS, drone data is also used.
Most papers in the SOC domain make use of Digital Elevation Data to estimate SOC but none uses AGB as a predictor.

\begin{table}[h]
\centering
\begin{adjustbox}{width=\textwidth,center=\textwidth}
\begin{tabular}{ | m{4em} | m{15em} | m{8em} | m{8em} | m{8em} | m{5em} | m{4em} | m{4em} |} 
\hline
\textbf{Literature} & \textbf{Description} & \textbf{ML Techniques} & \textbf{Vegetation Cover} & \textbf{Data Sources} & \textbf{Region/Year of Study} & \textbf{Use Digital Elevation} & \textbf{Use AGB as a predictor}\\
\hline
Zhou et al. \cite{zhou_2021} 
& Estimated SOC at different spatial resolution at 20m, 100m, 400m and 800m. Analysed multiple models with different combinations of satellite input and three ML techniques.
& SVM, BRT, RF; R$^2$ = 0.470 for prediction at 100m spatial resolution and with all available predictors
& Forest Ecosystems and cultivated land
& Sentinel 2, Sentinel 3, Landsat 8
& Switzerland, 2021 & No & No \\
\hline
Zhou et al. \cite{zhou_2020}
& Estimated SOC in Slovenia using LUCAS \citep{LUCAS_SOIL} dataset as ground truth. With Sentinel Satellites and DEM data as predictors, the paper analysed the effect of sets of predictors and ML algorithms has on SOC estimation
& BRT, RF, SVM; R$^2$ = 0.38 for BRT and RF technique with all available data as predictor.
& Forest Ecosystems and cultivated land
& Sentinel 1, Sentinel 2, Digital Elevation Map
& Slovenia and a small part of Austria and Italy, 2020
& Yes & No \\
\hline
Emadi et al. \cite{nortern_iran_2020_soc}
& Analysed and compared ML techniques on predicting and mapping SOC content. Found most important variables for predicting SOC. Improved SOC estimation results using ANN and DNN. 
& SVM, ANN, RF, XGBoost, DNN (R$^2$ = 0.65 for DNN)
& All types (Forests, cultivated land, residential land, seashore land)
& LandSat 8, MODIS, DEM, WorldClim \citep{worldclim}
& Mazandaran province, Northern Iran, 2020 
& Yes & No \\
\hline
Gholizadeh et al. \cite{SOC_cezch_2018}
& Studied SOC estimation and mapping through Satellite data, Airbourne data and field collected data. Focused on 4 specific agricultural farmland sites, predicted sand, slit and clay on top of SOC.
& SVM (RMSE = 0.12, 0.20, 0.09, 0.08 for the 4 sites respectively)
& Agricultural land
& Sentinel 2, Airbourne RS \citep{aircraft_RS}, DEM
& Přestavlky, Šardice, Nová Ves and Jičín in the Czech Republic, 2018
& Yes & No \\
\hline
\end{tabular}
\end{adjustbox}
\caption{Summary of state-of-the-art SOC estimation literature}

\label{table:soc_literature}
\end{table}

\subsection{Joint study of AGB and SOC as Total Carbon}
In remote sensing, AGB and SOC predictions are usually seen as two separate problems, however, there has long been research papers suggesting in order to understand the effect of climate change and the total carbon sink on earth, we need to study land as a terrestrial system. Early efforts provide carbon estimate in the vegetation and soils in the Great Britain \citep{carbon_gb_map_1995} based on land cover and allometric equations. Although the accuracy is limited and the carbon map produced is solely based on land cover estimations, the effort shows the research interest in providing information on AGB and SOC together. Scurlock and Hall \citep{global_carbon_grassland_2002} looked at the problem from a grassland perspective and introduced the concept of ``missing sink", which refers to the natural carbon sink that we have not been able to recognise before. The idea of grassland carbon sink is later extended \citep{Grassland_ECS_2019}, SOC, AGB, Grazing and Climate Change  are associated together and have a proven strong relationship between one another. Carlos et al. \citep{colombia_tcs} looked into the tropical forest landscape as an ecosystem to better understand the global carbon cycle and total carbon stock.
\newline \indent
Total Carbon Stock has also been analysed on the scale of a terrestrial ecosystem. Sothe et al. \cite{terrestrialcanada2021} analysed the total carbon stock in Canada by using different research sources. Although most sources either focuses on SOC and AGB separately, it is clear that both are required to provide useful information that can affect government level decision making and possibly the voluntary carbon market. Despite the abundance on separate research on SOC and AGB, there is a need to better understand total carbon through joint research on SOC and AGB. The reason behind is that terrestrial ecosystems are intertwined between AGB and SOC. The dynamics of carbon in terrestrial ecosystems are determined by processes such as respiration, combustion and decomposition \citep{india_markov_tc}.
\newline
Previous literature on total carbon estimation is summarised in table \ref{table:total-carbon-literature-summary}. Analysis in this domain is predominantly done using regression analysis \citep{Gebeyehu2019} \citep{total-1} \citep{total-2} \citep{Tibet_grassland_2008_regression}. Others \citep{multivariate_loess_plateau_2018} used ML techniques suggested in the AGB domain and SOC domain, but only random forest is explored. These research mainly focuses on grassland and forest ecosystems. While some of them used elevation data, none used satellite data from the Sentinel family (Sentinel 1,2,3) and LandSat 8 nor did any use SOC as a predictor. Some found the positive correlation between AGB and SOC \citep{total-2} \citep{Gebeyehu2019} but those only used field samples and LiDAR data which is not scalable to a larger study region.

\begin{table}
\centering
\begin{adjustbox}{width=1\textwidth,center=\textwidth}
\begin{tabular}{ | m{4em} | m{15em} | m{10em} | m{6em} | m{5em} | m{7em} | m{3em} | m{3em} | m{3em} |} 
\hline
\textbf{Literature} & \textbf{Description} & \textbf{ML Techniques} & \textbf{Vegetation Cover} & \textbf{Data Sources} & \textbf{Region/Year of Study} & \textbf{Use Digital Elevation} & \textbf{Use AGB as a predictor of SOC} & \textbf{Use SOC as a predictor of AGB} \\
\hline
Gebeyehu et al. \cite{Gebeyehu2019}
& Studied relationship between AGB/SOC and concluded from regression analysis that the significant positive correlation suggest AGB being a useful predictor of SOC.
& LR
& Forest Ecosystem
& Global Wood Density database, field samples
& Awi Zone, northwestern Ethiopia, 2019
& No & Yes & No \\
\hline
Wang et al. \cite{multivariate_loess_plateau_2018}
& Created multivariate RF model to estimated topsoil SOC and AGB, using meteorological factors, Satellite images and Digital Elevation. Discovered strong positive correlation between AGB and SOC in desert steppe and the steppe desert of rocky mountains. Provided evidence that AGB and air temperature should be given more attention in SOC prediction.
& RF (R $^2$ = 0.62, RMSE = 89.37 for AGB and R $^2$ = 0.72, RMSE = 3.99) 
& Grassland 
& MODIS, LandSat 5, ASTER (Elevation)
& Loess Plateau, China, 2017 & Yes & No & No\\
\hline
Vicharnakorn et al. \cite{total-1}
& First perform land classification, then developed AGB estimation model from field samples and Landsat Thematic Mapper (TM) image. Various bands were analysed with multiple regression analysis to study the correlation between AGB and RS bands. This is later put together with field measured SOC to present a total carbon estimation for the study area
& LR; Correlation in regression model between AGB and RVI/SAVI/SR is 0.931
& Forest Ecosystem
& Landsat Thematic Mapper (TM) image
& Savannakhet Province, Lao People’s Democratic Republic, 2014 
& No & No & No\\
\hline
Rasel \cite{total-2}
& Analysed AGB, elevation, bulk density and soil PH in the context of SOC. Found a positive correlation between SOC and elevation and AGB
& Linear Regression for AGB estimation, which is then used to study SOC content. Correlation of 0.79 between AGB/SOC and 0.84 between AGB/Elevation
& Forest Ecosystem
& LiDAR, DEM, AGB 
& Chitwan district, Nepal, 2013 & Yes & Yes & No \\
\hline
Yang et al. \cite{Tibet_grassland_2008_regression}
& Examined the relationship between AGB/SOCD and found strong positive correlation, suggesting plant production largely determine SOC content in alpine grassland. EVI derived from MODIS also has strong correlation between AGB and SOC Density, and is treated as a predictor variable for SOC estimation. 
& Regression Analysis (R$^2$ SOCD/AGB = 0.39)
& Alpine Grassland
& MODIS-EVI
& Qinghai-Tibetan Plateau, China, 2008 & No & No & No \\
\hline
Scurlock \& Hall \cite{global_carbon_grassland_2002}
& Discovered that grassland and savannas contribute to more 'missing sink' than previously anticipated, suggesting possible future research directions
& N/A
& Grassland
& Field measurements and Previous studies
& Global, 1997 & N/A & N/A & N/A\\
\hline
Milne \& Brown \cite{carbon_gb_map_1995}
& Created total carbon map for the Great Britain by combining previous studies on biomass partitioning, census of forests, ecological surveys of sample areas and RS land cover map. Suggesting early interest in the total carbon estimation domain combining SOC/AGB
& N/A
& General to the UK
& Past studies 
& Great Britain, 1995 & N/A & N/A & N/A \\
\hline
\end{tabular}
\end{adjustbox}
\caption{Summary of joint research on AGB/SOC estimation}
\label{table:total-carbon-literature-summary}
\end{table}

\subsection{State of the art and comparison}
State of the art models are used as baseline for comparison. The SOC baseline is trained on Sentinel 1A, Sentinel 2A (Including vegetation indices) and Digital Elevation Data. This is based on previous studies: Zhou et al. \cite{zhou_2020} used Sentinel 1/2 and DEM data, whereas Emadi et al. \cite{nortern_iran_2020_soc} used Digital Elevation Data, Gholizadeh et al. \cite{SOC_cezch_2018} used Sentinel 2 and Digital Elevation data. 
On the other hand, the AGB baseline model is based on Li et al. \cite{li_2020} which used LandSat 8 and Forest Inventory data as they contain information on multiple wavelengths and woodland classification which are essential to above ground biomass estimates. 
\newline

The experimentation involves a number of stages. To begin, we apply state-of-the-art models/input variables to our site in order to create a baseline model from which we can improve. The baseline models are then improved through feature engineering and an examination of the relationships between AGB and SOC. Although there has been research on the correlation between AGB and SOC, none has examined the possibility of using one as an input to predict the other. As a result, our investigation in AGB and SOC has two primary objectives:
\begin{enumerate}
\item Mix inputs that were previously thought to be useful only for AGB or SOC. For instance, given that digital elevation is known to be a good predictor for SOC \citep{zhou_2020}, we ask whether it also provides insight into AGB prediction.
\item Since there exist positive correlation between AGB and SOC \citep{total-2}, we attempt to improve existing models by including SOC/SOC Density to predict AGB, and use AGB to predict SOC.
\end{enumerate}

\section{Materials and methods}
\label{sec:materials_and_methods}

\subsection{Study Area}
\label{subsec:study_area}
\begin{figure}
    \centering
    \includegraphics[width=.8\linewidth]{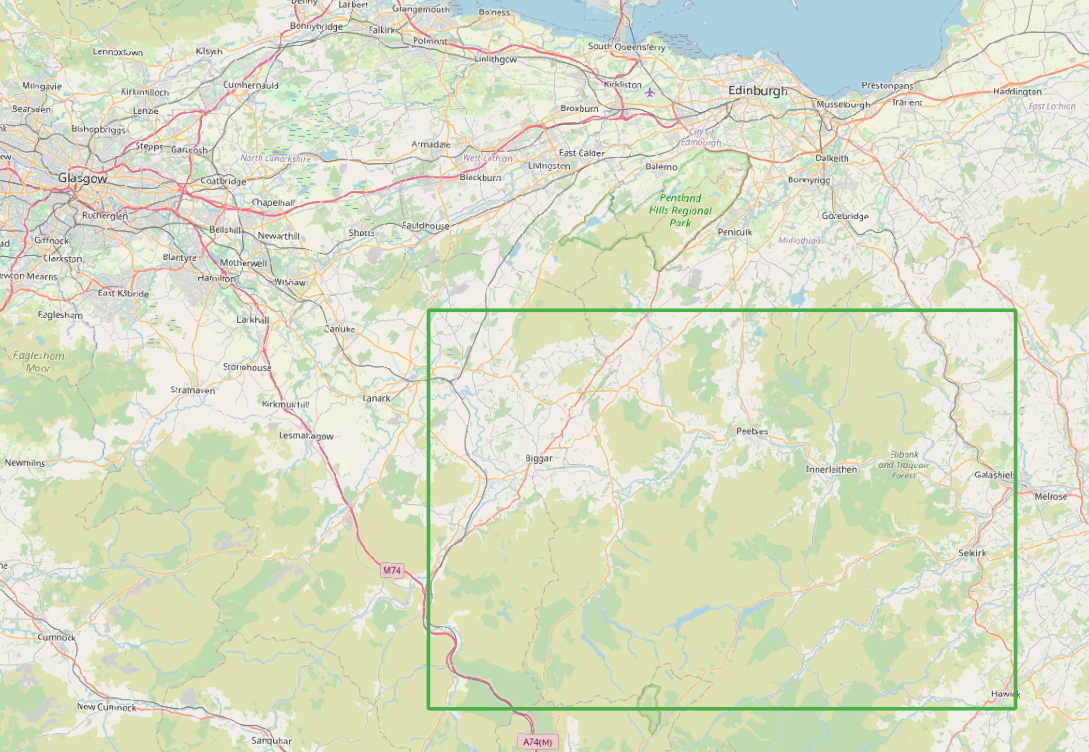}
    \caption{The study area on map}
    \label{fig:scotland-area}
\end{figure}

The study area covers a part of rural area in Scotland, United Kingdom. It is located east of Glasglow and south of Edinburgh between (Latitude, Longitude) = (55.754194, -3.703772) NW and (55.4075244°, -2.7696528°) SE. The study area is shown in figure \ref{fig:scotland-area} , it is covered by a mix of forests, grassland and other urban areas. From the National Forest Inventory Woodland Scotland data \citep{inventory_data_source}, the identified forest inventory is mainly covered by woodland (92.91\%), mixing some grassland (2.62\%) and urban areas (0.20\%).


\subsection{Data sources}
\subsubsection{Soil organic carbon}
A total of 25021 pre-processed data points representing Soil Organic Carbon at 0-30cm depth are obtained from Soil Grids 2.0 \citep{soil_grid}. Soil Grids 2.0 maps soil properties globally at a resolution of 250m, taking as input field soil samples from about 240000 locations worldwide. Soil samples in Soil Grids 2.0 are obtained from ISRIC World Soil Information Service (WoSIS), which provides global standardised soil profile data \citep{wosis_paper}. 

\subsubsection{Above ground biomass}
AGB data is obtained from the Global Above and Below Ground Biomass carbon density map \citep{Spawn2020}. The dataset \cite{agb_map_2010} is open sourced by NASA ORNL (Oak Ridge National Laboratory) featuring AGB at a resolution of 300m. The global map is compiled from published literature using a harmonization approach, matching maps of tundra, grassland and annual crops.

\subsection{Predictor variables}
The predictor variables used in this paper are Sentinel 1, Sentinel 2, Landsat 8, DEM derivatives and Scotland Forest Inventory Data. These variables are obtained from various sources and converted into raster data (300m) using QGIS 3.16.6 with Grass 7.8.5. All predictor variables, AGB and SOC were transferred to the OSGB 1936 / British National Grid projection geographic information system for analysis.

\subsubsection{Topographic variables}
DEM data (EU-DEM v1.1) at a resolution of 25m was obtained from the Cornipicus Land Portal \cite{dem_source}. It is an upgrade from EU-DEM v1.0, which is generated from SRTM and ASTER GDEM data, through further corrections and improvements. Four DEM derivatives were calculated using QGIS 3.16.6 and SAGA GIS software, these includes elevation, cathcment slope (CS), length-slope factor (LSF), topographic wetness index (TWI).

\subsubsection{Inventory variables}
Forest Inventory data is obtained from National Forest Inventory (NFI) \citep{inventory_data_source} developed by the English Forestry Commission. The NFI woodland map provides information on forest and woodland area with a minimum of 20\% canopy cover over 0.5 hectares. The vector data is one hot encoded to represent woodland and non woodland classification.

\subsubsection{Remote sensing variables and processing}
The remote sensing data for modelling included S1 and S2 images downloaded from ESA Copernicus Open Access Hub, and LandSat 8 images downloaded from ESGS Earth Explorer. Sentinel 1A data uses SAR (Synthetic Aperture Radar) and records backscatter. This study uses one image using the Interferometic Wide Swath (IW) acquisition mode \citep{s1_acquisition}. The polarisation is Vertical Transmit-Vertical Receive Polarisation (VV) and Vertical Transmit-Horizontal Receive Polarisation (VH), which measures volume scattering and rough surface scattering \citep{what_sar}. The image is taken on 5th May 2021, at cycle 230, orbit 52. Sentinel 2A \citep{s2a_user_guide} is a wide-swath and multi-spectral satellite. The Multi-spectral Instrument (MSI) samples 13 spectral bands at various resolutions with wavelengths from 442.4 to 2202.4 nanometers \citep{s2a_wavelen}. A cloudless Sentinel 2A (Level 1C product) image was captured on 10th October 2018.
LandSat 8 carries two sensors, Operational Land Imager (OLI) and the Thermal Infrared Sensor (TIRS). The two sensors provide global coverage at multiple spatial resolutions \citep{landsat8}. The LandSat image was captured on 6th May 2020.
\newline\indent
S1 SAR Data was pre-processed using the SNAP software: apply orbit file, radiometric callibration, speckle filtering (Lee filter 13x13) and terrain correction. To match the resolution of the AGB data, all images were downsampled to 300m using the nearest neighbour algorithm in QGIS 3.16.6. S2 data is processed using the sen2Cor processor which applies atmospheric correction and transform the data product from Level-1C (Top of atmosphere reflectance image) to Level-2A (Bottom of atmosphere reflectance image). Level-L1TP LandSat 8 images were preprocessed through the LandSat Product Generation System (LPGS), which used Ground Control Points (GCP) and DEM to callibrate radiometrically and orthorectify displacements.
\newline\indent
The backscatter coefficient from VH and VV polarization in S1 were calculated as environmental variables. Nine bands B2, B3, B4, B5, B6, B7, B8A, B11 and B12 were obtained from S2. Eleven bands L1 - L11 were extracted from LandSat 8. Additionally, three spectral indices were calculated from S2 Bands as predictors, these are reported to have strong correlation with AGB and SOC \citep{SOC_cezch_2018}. These indices are Nomalised Difference Vegetation Index (NDVI), Enhanced Vegetation Index (EVI) and Soil Adjusted Total Vegetation Index (SATVI), their formulas are as follows: 

\begin{align}
NDVI &= \frac{B8 - B4}{B8 + B4} \\
EVI  &= \frac{2.5 \cdot (B8 - B4)}{B8 + 6 \cdot B4 - 7.5 \cdot B2 + 1} \\
SATVI &= \frac{2(B11 - B4)}{B11 + B4 + 1} - \frac{1}{2}B12
\end{align}

\subsection{Modelling Techniques}
This paper used three machine learning techniques to estimate AGB and SOC content. The predictor variables and ground truth variables were first sampled from raw data source raster files and extracted in QGIS 3.16.6. Optimization were performed using grid search in sci-kit learn to tune hyper-parameters. The performance of the models with best parameters was then evaluated.

\subsubsection{Random Forest}
Random Forest \citep{randomforest201624} is an ensemble method which predicts through a set of classification and regression trees. These trees are created from a subset of training samples. The in-bag (About two thirds) samples are used to train trees and the remaining samples are regarded as out-of-the bag samples and used for evaluation. The error is estimated through out-of-bag (OOB) error. From the prediction of each tree then comprise the final output through voting or averaging.

\subsubsection{Boosted Regression Tree}
The Boosted Regression Tree model combines boosting techniques and decision tree algorithm for prediction. Boosting reduces overfitting by randomly select a subset of training data to fit new tree models. 
Compared to Random Forest models which use the bagging method, BRTs use the boosting method which weights input data in subsequent trees \citep{brt_explained}. Weighting in a way which poorly modelled data in previous trees has a higher probability of selection in the new tree. This improves the accuracy since the model will take into account the error of the previous tree to fit the current tree. 
\subsubsection{XGBoost}
Proposed by Chen et al. \citep{xgb_paper}, XGBoost is a very popular ML model upon its success in winning state-of-the-art performance in Kaggle ML competitions. XGBoost is an implementation of gradient boosted regression trees designed for performance and speed. It uses the second derivative of the objective function to accelerate convergences speed and reduces overfitting by adding a regularization term to the objective function. This results in a highly flexible and scalable model which handles sparse data with high convergence speed.

\subsection{Statistical Analysis}
Statistical analysis is performed to measure collinearity between predictor variables and AGB and SOC. This study used Gini coefficient and Pearson correlation from the SK Learn python package. Variables with high correlation (r $>=$ 0.9) and with high variance inflation factor (VIF $>=$ 10) were removed to form Model E and F. VIF is a ratio between the variance of the model of all variables and the variance of the model of one specific vairable. Equations \ref{pearson_equation} and \ref{vif_equation} shows the formula for VIF and Pearson Correlation used for our analysis.
\begin{equation}
\label{pearson_equation}
    r_j = \frac{\sum(x_i - \bar{x})(y_i-\bar{y})}{\sqrt{\sum(x_i - \bar{x})^2\sum(y_i-\bar{y})^2}}
\end{equation}
\begin{equation}
\label{vif_equation}
    VIF_j = \frac{1}{1 - r^2_j}
\end{equation}
The strategy was to eliminate one of the highly collinear variables indicated by VIF scores and Pearson correlation iteratively, until all selected variables have a VIF score of less than 10. This paper developed the BRT, RF, XGB models from sklearn ensemble methods ``Gradient Boosting Regressor", ``Random Forest Regressor", and xgboost 1.4.2 from python PyPI repository respectively.

\subsection{Model Performance Evaluation}
The AGB (Table \ref{agb_models}) and SOC (Table \ref{soc_models}) content models were built based on three machine learning techniques with different combinations of predictor variables and ground truth variables. A comprehensive list of data sources and their corresponding predictor variables is summarised in table \ref{data_sources_predictors}.
\newline\indent 
Model A was chosen from the state of the art SOC predictor variables mentioned in Zhou et al. \citep{zhou_2020}, which used S1, S2 and DEM as predictors. Model B was based on the state of the art AGB predictor variables suggested by Li et al. \citep{li_2020}, which used LandSat 8 and Inventory data. Model D combined the predictors from Model A and B, and was motivated by the following:
\begin{enumerate}
    \item Correlation between AGB and SOC suggests a relationship on their corresponding predictor variables \citep{total-2}.
    \item While DEM is a useful predictor for SOC, it can also have an effect on AGB as it affects air temperature, moisture and the above ground growth conditions for trees \citep{worldclim}. 
    \item Above ground vegetation plays an important role in soil condition and SOC content. Soil organic carbon are found to be richer in forest ecosystems \citep{soc_land_use_2014}, including inventory data in SOC prediction helps better locate forests ecosystems.
\end{enumerate}
Model E and model F was created after performing statistical analysis from Model D predictor variables on SOC and AGB. On top of statistical analysis, model G and model H explored the indirect relationship between AGB and SOC by including them as predictor variables to predict one another target variables. This paper used 5-fold cross validation to evaluate the performance of the models.
Three metrics were used to assess the model's performance. MAE and RMSE were used to quantify the difference in error between predictions and ground truth variables, whereas $R^2$ (Coefficient of Determination) was used to quantify how well the model accounts for the variability of input data around its mean. The formulas are demonstrated in the equations \ref{eq:rmse}, \ref{eq:mae} and \ref{eq:r2}. In general, a higher $R^2$ value and lower RMSE/MAE value indicate better estimation performance of the model. 
\begin{equation}
\label{eq:rmse}
    RMSE = \sqrt{\frac{1}{n} \sum^n_{i=1} (Y_i - X_i)^2}
\end{equation}
\begin{equation}
\label{eq:mae}
     MAE = \frac{1}{n} \sum^n_{i=1} | Y_i - X_i |
\end{equation}
\begin{equation}
\label{eq:r2}
R^2 = \frac{\sum^n_{i=1} (Y_i - \bar{X_i})^2}{\sum^n_{i=1} (X_i - \bar{X_i})^2}
\end{equation}

\begin{table}
\begin{threeparttable}
\caption{Predicting AGB with different combinations of Sentinel 1, Sentinel 2, LandSat 8, DEM derivatives, forest inventory, AGB and SOC data} \label{agb_models}
\begin{tabular}{ m{3em} c c } \toprule
No. & Model  & Data Sources \\
\\ \midrule
i & Model A & S1, S2 and DEM\\
ii & Model B & L8 and Inventory Data\\
iii & Model C & S1, DEM and Inventory Data\\
iv & Model D & S1, S2, L8, DEM and Inventory Data\\ 
v & Model F & S1, S2 (Band  4, 8A), NDVI, DEM, \\
&& L8 (Band 5,6,8,9) and Inventory Data \\
vi & Model H & S1, S2 (Excluding Band 2 and 3), DEM (CS, Elevation), \\ 
&& L8 (Band 5-7, 10, 11), Inventory Data, SOC, SOCD \tnote{a} \\
\bottomrule
\end{tabular}
\begin{tablenotes}
\item[a] Soil Carbon Density (SOCD)
\end{tablenotes}
\end{threeparttable}
\end{table}

\begin{table} 
\begin{threeparttable}
\caption{Predicting SOC with different combinations of Sentinel 1, Sentinel 2, LandSat 8, DEM derivatives, forest inventory, AGB and SOC data}\label{soc_models}
\begin{tabular}{ m{3em} c c } \toprule
No. & Model & Data Sources \\
\\ \midrule
i & Model A & S1, S2 and DEM\\
ii & Model B & L8 and Inventory Data\\
iii & Model C & S1, DEM and Inventory Data\\
iv & Model D & S1, S2, L8, DEM and Inventory Data\\ 
v & Model E & S1, S2 (Band 2, 8A), EVI, DEM Derivatives,  \\
&& LandSat 8 (Band 4,5,6,10), Inventory Data \\
vi & Model G & S1, S2, DEM, AGB \\
\bottomrule
\end{tabular}
\end{threeparttable}
\end{table}

\begin{table} 
\begin{threeparttable}
\caption{Data sources and their corresponding predictors} \label{data_sources_predictors}
\begin{tabular}{ c  c } \toprule
\textbf{Data Source} & \textbf{Environmental Variables} \\
\\ \midrule
Sentinel 1 (S1) & VH, VV \\
Sentinel 2 (S2) & Band 2-7, 8A, 11, 12, EVI, NDVI, SATVI \\
DEM Derivatives & Elevation, CS \tnote{a} , LSF \tnote{b} , TWI \tnote{c} \\
LandSat 8 (L8) & Band 1 - 11 \\
Inventory Data & Woodland category \\
\bottomrule
\end{tabular}
\begin{tablenotes}
\item[a] Catchment Slope (CS) 
\item[b] Length Slope Factor (LSF)
\item[c] Tropical Wetness Index (TWI)
\end{tablenotes}
\end{threeparttable}
\end{table}

\section{Results}
\subsection{Analysis of results}
The SOC content is converted using natural logarithm for all prediction models, which reduces the variability of data for more stable training. Through collinenarity analysis, there exists high collinearity and correlation in S2 and LandSat 8 variables, all collinear variables with VIF score $>=$ 10 were removed and reflected in Model E,F,G and H.

\subsection{Evaluation and Comparison between models}
\begin{figure}
    \centering
    \begin{subfigure}{.9\textwidth}
        \centering
        \includegraphics[width=\linewidth]{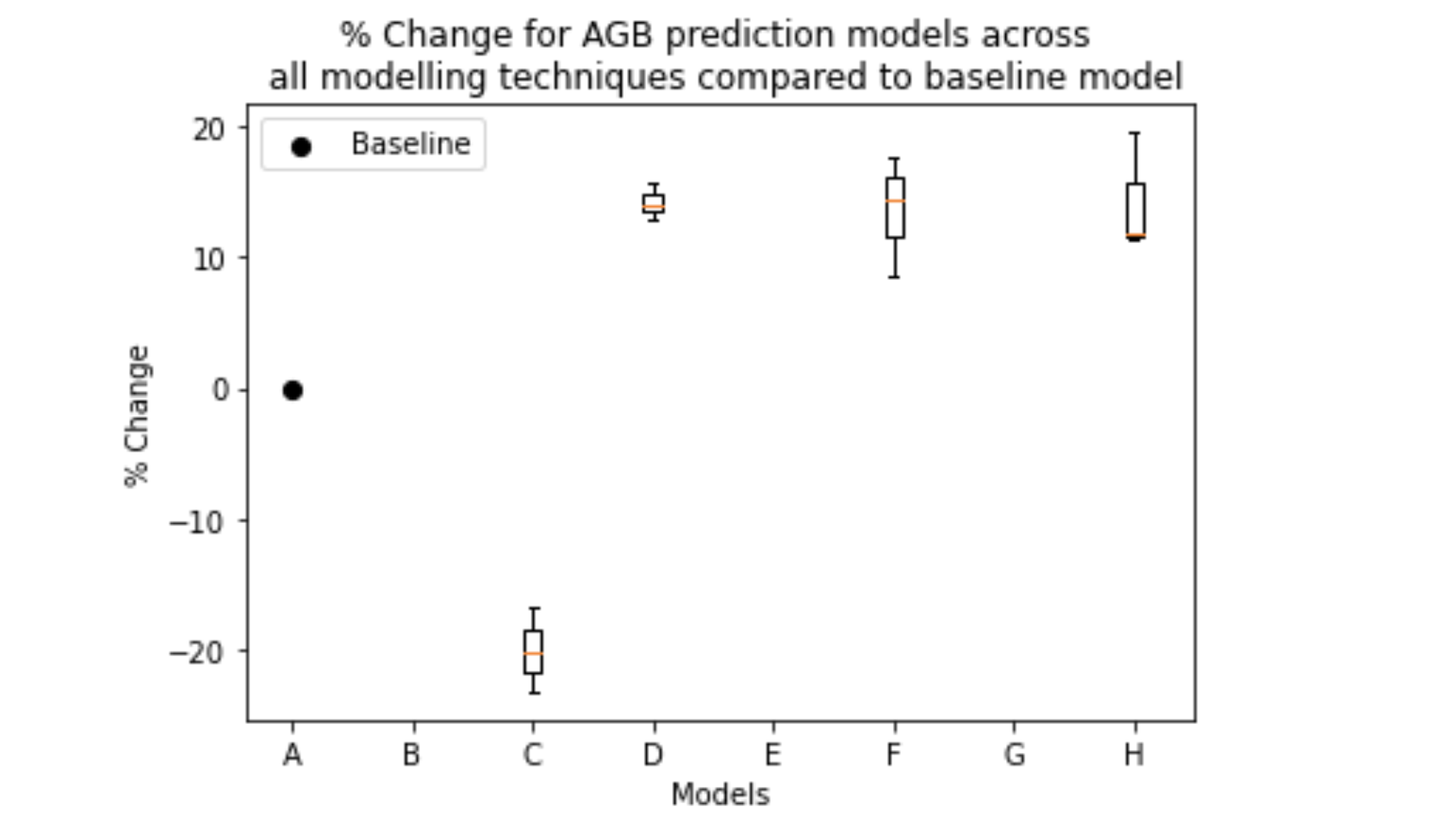}
        \caption{AGB Prediction percentage difference compared to baseline}
    \end{subfigure}%
    
    \begin{subfigure}{.9\textwidth}
        \centering
        \includegraphics[width=\linewidth]{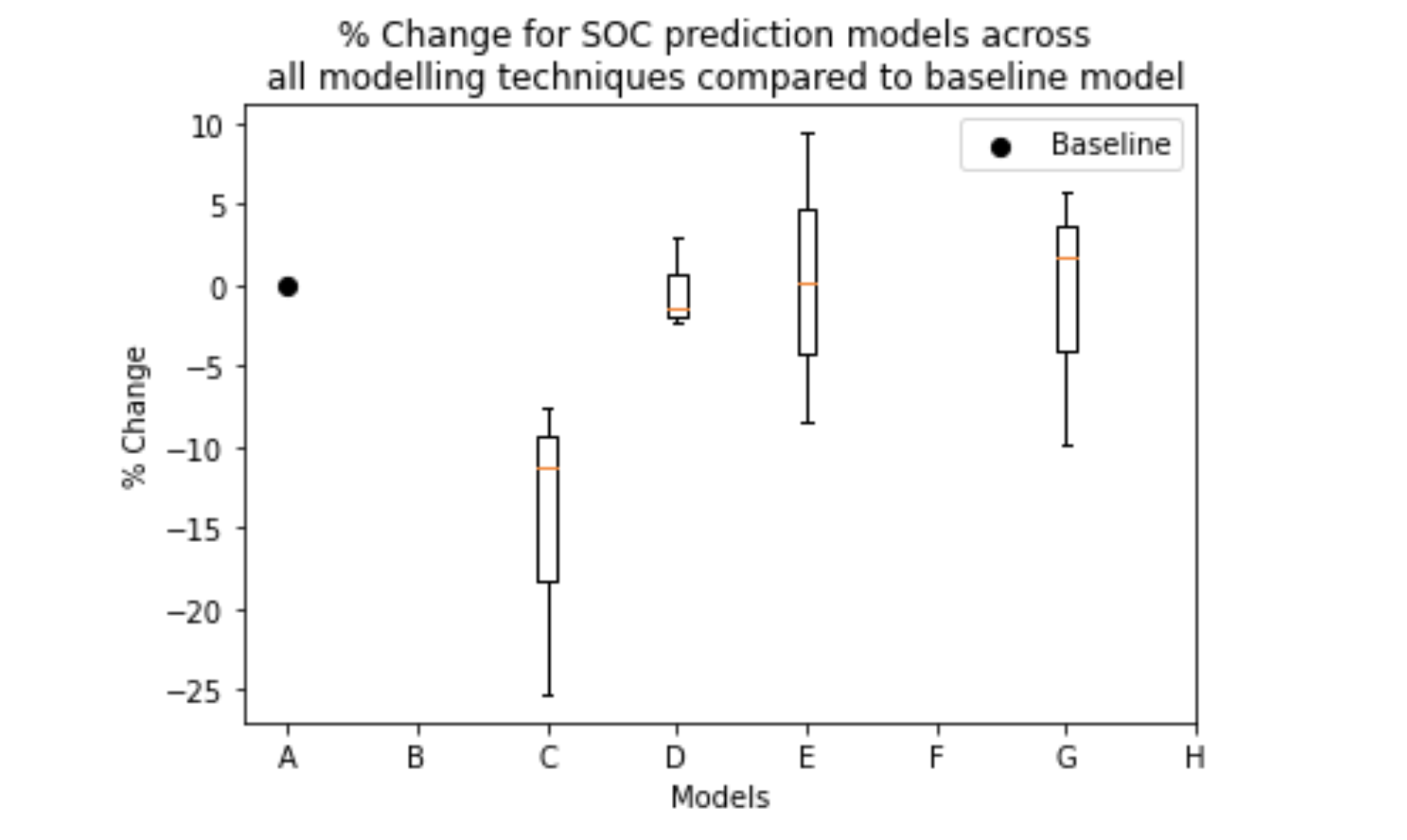}
        \caption{SOC Prediction percentage difference compared to baseline}
    \end{subfigure}%
    \caption{Percentage difference of different models compared to baseline models across all machine learning techniques}
    \label{fig:Percentage Diff for all models}
\end{figure}

This paper built eight models using Sentinel 1, Sentinel 2, LandSat 8, DEM derivatives, Forest Inventory data, AGB and SOC as predictors: Model A and model B represent the state of the art baseline models representing S1, S2, DEM to predict SOC and LandSat 8, Inventory Data to predict AGB; Model D represented all environmental variables; Model E and model F are derived from performing statistical analysis on model D and AGB and SOC, representing selected environmental variables; Model G and model H jointly study AGB and SOC, which used AGB to predict SOC and SOC to predict AGB in addition to the selected environmental variables.
\newline\indent The performances for Boosted Regression Tree, Random Forest and XGBoost based on these models are shown in table \ref{performances-01}. Through comparative analysis on prediction accuracy, it is observed that the different combinations of predictor variables and the choice of machine learning technique significantly affect AGB and SOC prediction performances. \textbf{In AGB predictions} using BRT and RF, Model B ($R^2$ = 0.5016 vs $R^2$ = 0.4958), Model D ($R^2$ = 0.5829 vs $R^2$ = 0.5734) and Model F ($R^2$ = 0.5898 vs $R^2$ = 0.5674) is better predicted by BRT, whereas Model H ($R^2$ = 0.5604 vs $R^2$ = 0.5925) is better predicted by RF. BRT and XGB have similar performances in Model D ($R^2$ = 0.5829), and XGB performed better in Model B (BRT $R^2$ = 0.5016 vs XGB $R^2$ = 0.5161) and Model H (BRT $R^2$ = 0.5604 vs XGB $R^2$ = 0.5750). \textbf{In SOC predictions}, the best results in Model A ($R^2$ = 0.7443) and Model D ($R^2$ = 0.7264) came from BRT, the best result for Model E ($R^2$ = 0.7518) came from XGB and the best result for Model G ($R^2$ = 0.7705) came from RF. Overall, the three machine learning techniques had varying performances with one better than the other in specific models. Figure \ref{fig:Percentage Diff for all models} shows box plots illustrating the \% increase in performance across all machine learning techniques for each model compared to the baseline. While different modelling techniques suits different environmental variables in predicting AGB and SOC, we can see consistent performance increase in AGB performance in model D, F and H, whereas the improvement for SOC is specific to ML modelling techniques and more research is required to prove consistent improvements.
\newline \indent
Throughout the different types of predictors, using S1, S2 and DEM improves AGB prediction by a significant amount. This is reflected when comparing Model B and Model D in all three machine learning techniques: BRT (From $R^2$ = 0.5016 to $R^2$ = 0.5829), RF (From $R^2$ = 0.4958 to $R^2$ = 0.5734), XGB (From $R^2$ = 0.5161 to $R^2$ = 0.5829). For SOC, when comparing Model A and Model D, introducing LandSat 8 and Inventory Data improves performance when modelling with XGB (From $R^2$ = 0.6871 to $R^2$ = 0.7070). However, there is a slight decrease in performance in BRT (From $R^2$ = 0.7443 to $R^2$ = 0.7264) and RF (From $R^2$ = 0.7289 to $R^2$ = 0.7185) models. Using SOC as a predictor for AGB (Comparing Model D and Model H) improves performance in RF (From $R^2$ = 0.5734 to $R^2$ = 0.5925) and using AGB as a predictor for SOC (Comparing Model D and Model G) significantly improves RF performances (From $R^2$ = 0.7295 to $R^2$ = 0.7705).
\newline \indent
Reflected in the results and figure \ref{fig:Percentage Diff for all models}, combining all environmental variables (S1, S2, LandSat 8, DEM, Inventory Data) improves AGB prediction performance by an average of 14.9\% (Comparing Model B and Model D, BRT improved by 16.2\%; RF improved by 15.7\%; XGB improved by 12.9\%). This demonstrates that using environmental variables previously known to be good predictors for other target variables can be critical to improving modelling performances. Through applying SOC as a predictor for AGB (Comparing Model A and Model H), the performance in RF improved by 19.5\% compared to baseline, while on average an increase of 14.2\% (BRT improved by 11.7\%; RF improved by 19.5\%, XGB improved by 11.4\%).

\begin{table*} 
\begin{threeparttable} 
\caption{Prediction accuracy of AGB and SOC with different combinations of predictors. The most accurate results are shown in bold.} \label{performances-01}
\begin{tabular}{ c c c c c c c c } \toprule
Modelling technique & Model & AGB &&& SOC &&  \\
&& RMSE & MAE & $R^2$ & RMSE & MAE & $R^2$ \\
\\ \midrule
BRT & Model A & - & - & - & 0.3140 & 0.0968 & 0.7443 \\
& Model B & 173.4170 & 108.7244 & 0.5016 & - & - & - \\
& Model C & 186.2773 & 120.1271 & 0.4180 & 0.3045 & 0.0961 & 0.6887 \\
& Model D & 158.3030 & 100.4707 & 0.5829 & 0.3172  & 0.0973 &  0.7264 \\
&Model E & - & - & - & 0.3792  & 0.1064 &   0.6812\\
& Model F & \textbf{162.8238} & \textbf{103.5530} &  \textbf{0.5898} & - & - & - \\
& Model G & - & - & - & 0.3490 & 0.1038 & 0.6717 \\
& Model H & 163.8379 & 104.0099 & 0.5604 & - & - & -
\\
RF & Model A & - & - & - & 0.2944 & 0.0928 & 0.7289\\
& Model B & 178.5750 & 114.1506 & 0.4958 & - & - & - \\
& Model C & 190.9638 & 124.6872 & 0.4128 & 0.4021 & 0.1141 & 0.5447 \\
& Model D & 161.0494 & 102.9460 & 0.5734 & 0.3398 & 0.0964  & 0.7185 \\
& Model E & - & - & - & 0.3151 & 0.1064 &  0.7295\\
& Model F & 159.1182 & 101.0034 &  0.5674 & - & - & - \\
& Model G & - & - & - & \textbf{0.3075} & \textbf{0.0967} & \textbf{0.7705} \\
& Model H & \textbf{158.6507 }& \textbf{101.7742} & \textbf{0.5925} & - & - & -
\\
XGB & Model A & - & - & - & 0.3414 & 0.1129 & 0.6871\\
& Model B & 168.8985 & 107.0769 & 0.5161 & - & - & -\\
& Model C & 187.8227 & 119.4395 & 0.3965 & 0.3836 & 0.1212 & 0.6100 \\
& Model D & \textbf{159.7902} & \textbf{100.2105} & \textbf{0.5829} & 0.3450 & 0.1131  & 0.7070 \\
& Model E & - & - & - & \textbf{0.3239} & \textbf{0.1107} &  \textbf{0.7518} \\
& Model F & 162.5048 & 101.2534 & 0.5604 & - & - & - \\
& Model G & - & - & - & 0.3620 & 0.1158 & 0.6753 \\
& Model H & 160.8522 & 100.2997 & 0.5750 & - & - & - \\

\bottomrule
\end{tabular}
\end{threeparttable}
\end{table*}

\subsection{Feature Importance of predictors}
\begin{figure}
    \centering
    \begin{subfigure}{.5\textwidth}
        \centering
        \includegraphics[width=\linewidth]{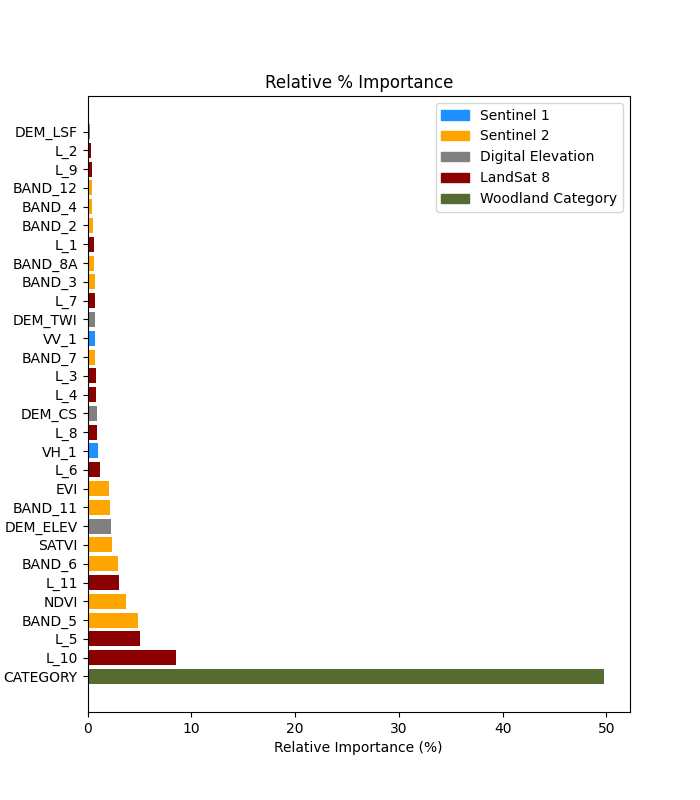}
        \caption{AGB BRT Model D}
    \end{subfigure}%
    \begin{subfigure}{.5\textwidth}
        \centering
        \includegraphics[width=\linewidth]{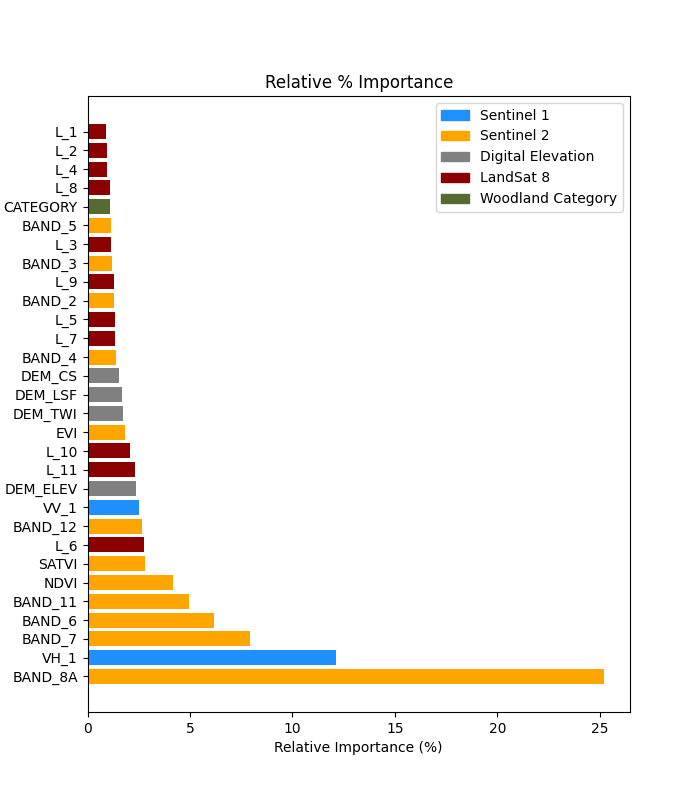}
        \caption{SOC XGB Model D}
    \end{subfigure}%

    \begin{subfigure}{.5\textwidth}
        \centering
        \includegraphics[width=\linewidth]{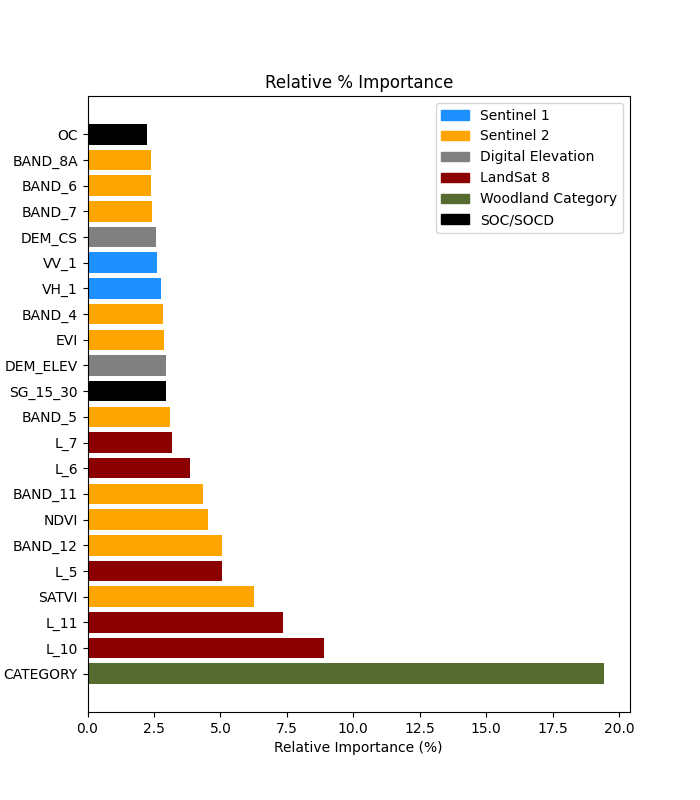}
        \caption{AGB RF Model H}
    \end{subfigure}%
    \begin{subfigure}{.5\textwidth}
        \centering
        \includegraphics[width=\linewidth]{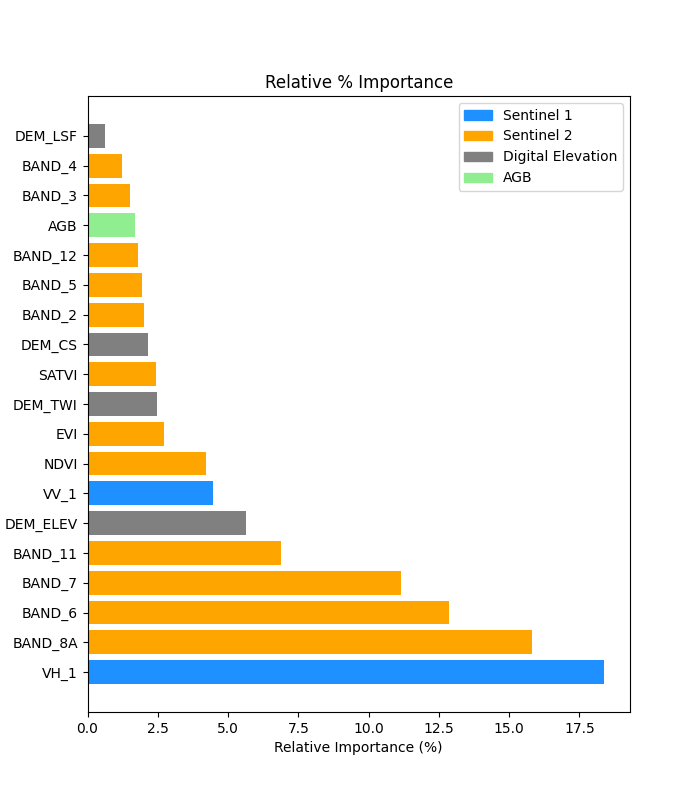}
        \caption{SOC RF Model G}
    \end{subfigure}%
    \caption{Feature Importance in Models}
    \label{fig:relative_importances}
\end{figure}

For AGB and SOC mapping with model D, model H and model G, the percentage relative importance of predictor variables are shown in figure \ref{fig:relative_importances}. Overall, BRT and XGB models depends heavily in one or a few predictors while RF models allow importance spreading across a wider range of variables. The AGB model predictions are heavily influenced by Inventory data, which is to be expected given that AGB is predominantly found in woodland areas. Notable is the fact that both Sentinel 2 and DEM derivatives contribute to the predictive power of AGB BRT Model D. This evidence substantiates our claim that combining SOC and AGB predictors improves AGB model estimation. In SOC models, we can see that Band 8A has the greatest impact on prediction, while Sentinel 1 data and digital elevation also play a role. 
In model G, although not being the most important factor, AGB still plays a role in SOC estimation and its importance is comparable to that of Sentinel 2 Bands (2-5). This explains the slight improvement in SOC estimation from RF Model D ($R^2$ = 0.7185, MAE = 0.0964, RMSE = 0.3398) to RF Model G ($R^2$ = 0.7705, MAE = 0.0967, RMSE = 0.3075), although AGB has some influence, it is not the most important variables in predicting SOC. The model prediction is still dominated by Sentinel 1 data and Band 8A from Sentinel 2.
For Model H, inventory data still has a very large influence in the model as expected, followed by Sentinel 2 and Landsat 8 data. It is interesting to see that Soil Carbon Density is now more important than Digital Elevation and Sentinel 1 data, verifying the positive correlation between AGB and SOC discovered in previous literature \citep{total-2} \citep{Gebeyehu2019}. Despite the correlation discovered, Rasel \cite{total-2} only experimented the possible effect of AGB on SOC Estimation. Whereas we have now proved the other way round as well - using SOC and SOC Density improves AGB Estimation performance.

\newpage

\subsection{Spatial characteristics of AGB and SOC maps}
\begin{figure}
    \centering
    \begin{subfigure}{\textwidth}
        \centering
        \includegraphics[width=\linewidth]{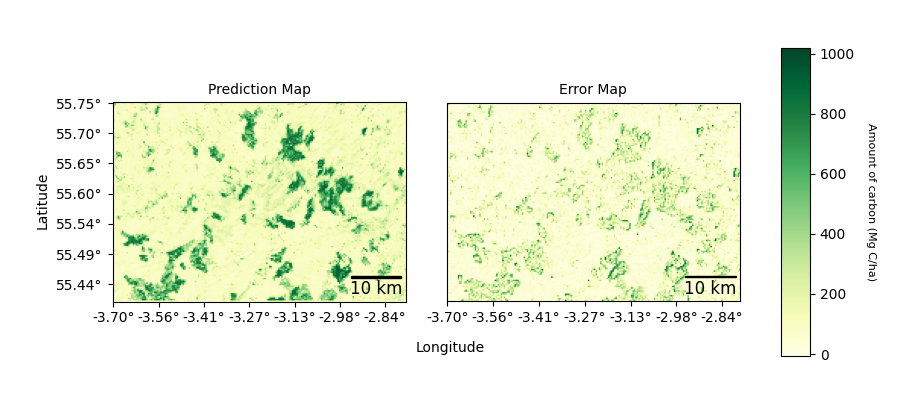}
    \end{subfigure}%
    \caption{AGB Carbon Prediction and Error Maps}
    \label{fig:agb_prediction_maps}
\end{figure}

\begin{figure}    
    \begin{subfigure}{\textwidth}
        \centering
        \includegraphics[width=\linewidth]{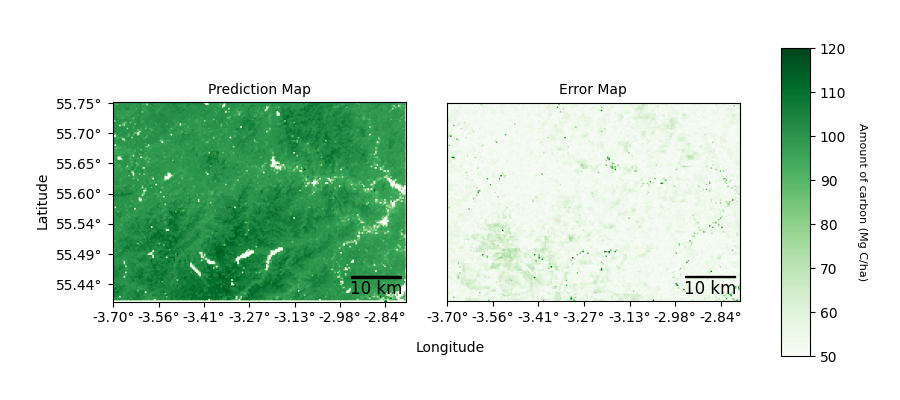}
    \end{subfigure}%
    \caption{SOC Carbon Prediction and Error Maps}
    \label{fig:soc_prediction_maps}
\end{figure}

\begin{figure}
    \centering
    \begin{subfigure}{\textwidth}
        \centering
        \includegraphics[width=\linewidth]{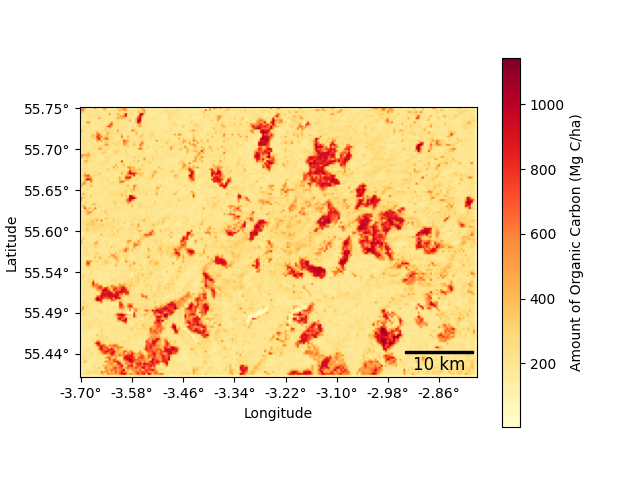}
        \caption{Total Carbon Prediction}
    \end{subfigure}%
    
    \begin{subfigure}{\textwidth}
        \centering
        \includegraphics[width=\linewidth]{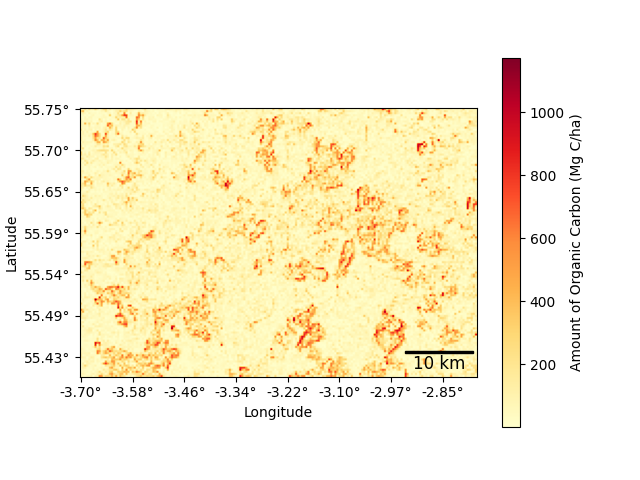}
        \caption{Total Carbon error}
    \end{subfigure}%
    
    \caption{Carbon Prediction and Error Maps for AGB and SOC}
    \label{fig:total_carbon_maps}
\end{figure}

\begin{figure}
    \centering
    \begin{subfigure}{.9\textwidth}
        \centering
        \includegraphics[width=\linewidth]{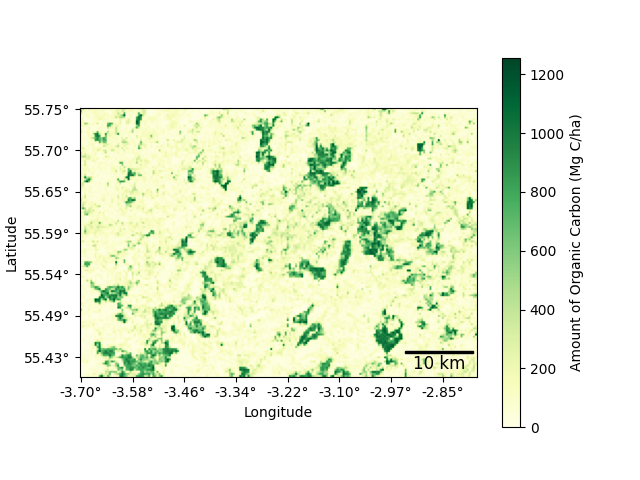}
        \caption{AGB Truth}
    \end{subfigure}%
    
    \begin{subfigure}{.9\textwidth}
        \centering
        \includegraphics[width=\linewidth]{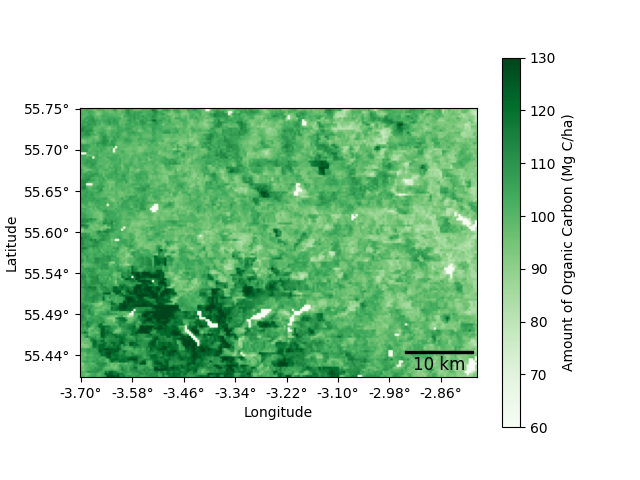}
        \caption{SOC Truth}
    \end{subfigure}%
    
    \begin{subfigure}{.9\textwidth}
        \centering
        \includegraphics[width=\linewidth]{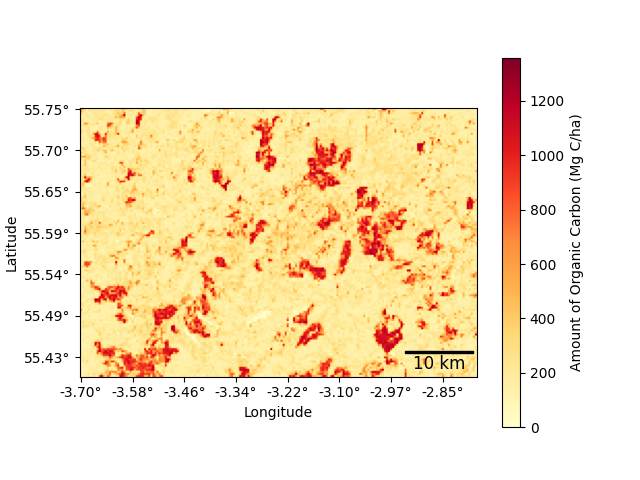}
        \caption{Total Carbon Truth}
    \end{subfigure}%
    \caption{Carbon Maps for AGB, SOC and Total Carbon ground truth}
    \label{fig:truth_maps}
\end{figure}

Carbon maps for AGB (Figure \ref{fig:agb_prediction_maps}) and SOC (Figure \ref{fig:soc_prediction_maps}) are obtained from model H and Model G predictions respectively. The total carbon maps in figure \ref{fig:total_carbon_maps} is generated from adding carbon predictions in both maps together. The error maps are created by the absolute difference between the predictions and ground truth carbon content. This can be compared against the ground truth map in figure \ref{fig:truth_maps}.

\section{Discussion}
\subsection{Performance of prediction models using Sentinel 1, Sentinel 2, LandSat 8, DEM and Forest Inventory Data}
\indent \textbf{Slight difference in SOC union models compared to baseline models:}
For the SOC models, Model A (baseline) and D (union model) performs similarly across all three modelling techniques, BRT shows a difference of ($R^2$ = 0.0179, MAE = 0.0005, RMSE = 0.0032), RF shows a difference of ($R^2$ = 0.0104, MAE = 0.0036, RMSE = 0.0454), XGB shows a difference of ($R^2$ = 0.0199, MAE = 0.0002, RMSE = 0.0036). The slight change in performance suggest predictors used in AGB estimation is not very useful to predict SOC. This can be explained by the fact that forest inventory data only identify forest areas \citep{inventory_data_source} but did not take into account the fact that soil organic carbon is also abundant in other land covers such as agricultural land. 
\newline\indent \textbf{VIF collinear variable removal improves performance: } 
Lombardo et al. \cite{lombardo2018}, suggested removing one of two or more collinear variables iteratively to avoid multicollinearity. Following this method, we created Model E (predicting SOC) which improved the XGB technique for SOC from the ($R^2$ = 0.6871, MAE = 0.1129, RMSE = 0.3414) baseline to ($R^2$ = 0.7518, MAE = 0.1107, RMSE = 0.3239) and Model F (predicting AGB) which improved the BRT technique from  The BRT technique for AGB also improved from ($R^2$ = 0.5829, MAE = 100.4707, RMSE = 158.3030) to ($R^2$ = 0.5898, MAE = 103.5530, RMSE = 162.8238)
On the contrary, if we remove all collinear variables (Model C) at once, the predictive power decreases for all modelling techniques. This is because removing multiple collinear variables simultaneously has the unintended consequence of removing information that is not highly collinear with the remaining variables. It is important to note that, although not experimented in this paper, the feature selection process can also leverage domain knowledge from SOC or AGB experts which can complement these ML techniques.
\newline\indent\textbf{Using Digital Elevation Data in AGB estimation models significantly improves performance:}
This is one of the major findings as there is a significant improvement in AGB model performances after including variables previously used for SOC model predictions. Zhou et al. \cite{zhou_2020} used Sentinel 1, Sentinel 2 and DEM data to predict SOC while Li et al. \cite{li_2020} used Sentinel 1 and LandSat 8 to predict AGB. Using predictors previously used in SOC prediction improves the AGB model by a significant 14.9\% across all ML techniques (Shown in table \ref{performances-01} and illustrated in figure \ref{fig:Percentage Diff for all models}). This indicates that Sentinel 2 and DEM data contain useful information for predicting AGB. There is no prior attempt in any literature to use Digital Elevation to estimate AGB. It demonstrates that factors associated with SOC may have an effect on predicting AGB.

\subsection{Spatial Characteristics of prediction maps}
From the total carbon and error maps, most prediction error come from above ground biomass concentrated regions, while we are very successful predicting the locations of high carbon content regions, the estimation in these regions still require more attention. There are two ways to mitigate this problem and improve our carbon map performance. 
\begin{enumerate}
    \item \textbf{Higher resolution study at specific regions of interest:} We encounter noisy data when attempting to map the entire region which consists of a mix of land use. If we can perform segmentation and target regions with high carbon content, then we can eliminate unnecessary noise and obtain better results. 
    \item \textbf{Remove area that is impossible to have above ground biomass:} While this might not be the case for SOC, it is possible to identify areas with no above ground biomass and eliminate those regions from our study. For instance, it can be clearly identified that roads and urban areas has no above ground biomass value. We can set the AGB values and the predictor values for those regions to 0. This is another way to remove noise such that our model can focus on predicting the highly carbon concentrated regions.
\end{enumerate}

\subsection{Novel Discoveries}

Table \ref{novel-performances} extracts the results for our joint study models. The random forest model beats the state-of-the-art result by 19.5\% in AGB estimation and by 14.2\% on average across all ML techniques (Comparison between RF Model B and Model H). This is consistent with the observation that there is a direct positive relationship between AGB and SOC \citep{total-2}. We were able to verify the correlation between AGB and SOC despite our study area consists of a mix of forest and agriculture land. This is expected to be more prominent if we restrict our study area to only forest areas \citep{Gebeyehu2019}.
We have demonstrated that using SOC/SOCD to predict AGB improves model results. Thus, joint study between AGB and SOC is a crucial direction for future research in the domain of total carbon estimation.

\begin{table} 
\begin{adjustbox}{width=1\textwidth,center=\textwidth}
\begin{threeparttable} 
\caption{Prediction accuracy of AGB and SOC for novel models G, H} \label{novel-performances}
\begin{tabular}{ c c c c c c } \toprule
Modelling technique & Model & Target & RMSE & MAE & $R^2$ \\
\\ \midrule
BRT & Model G & SOC & 0.3490 & 0.1038 & 0.6717 \\
& Model H & AGB & 163.8379 & 104.0099 & 0.5604 \\
RF & \textbf{Model G} & \textbf{SOC} & \textbf{0.3075} & \textbf{0.0967} & \textbf{0.7705} \\ 
& \textbf{Model H} & \textbf{AGB} & \textbf{158.6507 }& \textbf{101.7742} & \textbf{0.5925} \\ 
XGB & Model G & SOC & 0.3620 & 0.1158 & 0.6753 \\ 
& Model H & AGB & 160.8522 & 100.2997 & 0.5750 \\ 

\bottomrule
\end{tabular}
\end{threeparttable}
\end{adjustbox}
\end{table}

\subsubsection{Digital Elevation as predictor improves AGB estimation}
Through the experiment of mixing AGB and SOC predictors, we observed a significant increase in performance in AGB estimation through the use of Digital Elevation Map as a predictor. With an average increase of 13.53\% across all three ML techniques, we discovered a way to leverage the relationship between AGB and SOC to improve machine learning model results. This insight help future studies in the total carbon domain to identify the most important predictors for carbon estimation models.   

\subsubsection{SOC and SOC Density are good predictors for AGB models}
We experimented using SOC and SOC Density as predictors for AGB estimation models, the best performing machine learning technique increase performance from $R^2$ = 0.5829 in RF Model D to $R^2$ = 0.7705 in RF Model H. 

\subsubsection{Using AGB as a predictor for SOC improves performance for certain ML techniques}
We discovered the indirect relationship between AGB and SOC, upon using AGB as a predictor variable, we improved model performance from $R^2$ = 0.7185 in RF Model D to $R^2$= 0.7705 in RF Model G. However, when taking into account other ML techniques, there is no improvement on average, the improvement is therefore ML technique specific and more research is required. On the other hand, upon performing feature importance analysis, we discovered that AGB has certain importance in the SOC estimation model.

\section{Conclusion}
In this work, we proposed a general methodology to estimate total carbon content in an AFOLU area of Scotland. There are two novel experiments conducted that contribute to the remote sensing carbon estimation domain: \textbf{i) Create a union predictor model that consists of predictors from the SOC and AGB carbon estimation domain.} \textbf{ii) Explore the indirect relationship between SOC and AGB and improved carbon estimation performance through the use of target variables as predictors.}
The experimentation results suggest that joint study of AGB and SOC is important for carbon estimation as biomass and soil continuously exchange carbon in terrestrial ecosystems. Through feature engineering and the two novel experiments we conducted, we improved the state-of-the-art AGB estimation by 14.2\% on average across all ML modelling techniques discussed.

\begin{acknowledgement}
We thank you \textbf{Dr Thomas Lancaster} for his advice on project direction and possible improvements. \textbf{Harry Grocott and Rob Godfrey from Treeconomy} for their time discussing useful data sources and sharing their expertise in remote sensing. \textbf{Engineering Change} for their input with innovative ideas and ideology of an application to showcase carbon maps. Dr Pedro Baiz would like to thank the \textbf{Royal Society} for grant ERR\textbackslash19\textbackslash104 and \textbf{Prof J. McCann} for their support.
\end{acknowledgement}

\section*{Source Code}
\small \url{https://github.com/TerrenceCKCHAN/Carbon-Trading-Verfication}


\bibliography{example}

\end{document}